\documentclass[letter,aps,amsmath,amssymb,10pt,onecolumn]{revtex4}

\usepackage{graphicx}
\usepackage{graphics,color}
\begin{document}
\vspace {2.0 in}

%\documentclass[draft]{ws-procs9x6}
%\documentclass{ws-procs9x6}
%\usepackage{subfigure}
%begin{document}

\title{COMPLEX COOPERATIVE BEHAVIOUR IN RANGE-FREE FRUSTRATED MANY-BODY SYSTEMS}

\author{David Sherrington}
\affiliation{Rudolf Peierls Centre for Theoretical Physics, University of Oxford,
1 Keble Road, Oxford OX1 3NP, UK.}

\date{19 August 2008}

\begin{abstract}

A brief introduction and overview is given of the complexity that 
is possible and the challenges its study poses in many-body systems
in which spatial dimension is irrelevant and naively one might have 
expected trivial behaviour.

\end{abstract}

\maketitle

\section{Introduction}

This paper is concerned with many body systems and their cooperative 
behaviour; in particular when that behaviour is complex and 
hard to anticipate from the microscopics, even qualitatively 
and even when the 
systems are made up of simple individual units 
with simple inter-unit interactions. 

`Range-free' (or `infinite-ranged') refers to situations
where the
interactions are not dependent on the 
physical separations of individual units, and hence neither on 
the dimensionality nor on the structure of the embedding space. Such 
systems are also often referred to as `mean-field', since 
one can often show (and usually believes) that their behaviour 
in the thermodynamic limit ($N \to \infty$ units) is  identical 
to that of an appropriate mean-field approximation to a short-range system. 

`Frustration' refers to incompatability between different microscopic 
ordering tendencies.

Self-consistent mean-field theories do have the ability to 
describe spontaneous symmetry breaking and phase transitions and they
have played an important role in statistical 
physics. However as pure systems, without quenched Hamiltonian disorder
or out-of-equilibrium self-induced disorder, 
they do not exhibit the interesting non-simple dimension-dependent but 
details-independent (universal) critical behaviour 
whose study drove much of 
the interest of statistical mechanics in the seventies and 
eighties \cite{Cardy}. For this reason `mean-field' used to be 
interpreted as fairly trivial. 

On the other hand, with quenched disorder and frustration in 
their interactions range-free many-body systems can, and regularly 
do, exhibit behaviour that is complex and rich. This paper represents 
a brief introduction to and partial overview of such systems.

\section{General structure and features}

The general class of systems we consider can be summarized 
as characterised by schematic `control functions' of the form
$H(\{J_{ij...k}\},\{S_{i}\},X)$
where \newline
(i) in thermodynamics (statics) the $\{S\}$ are the variables 
and the $\{J\}$ are quenched (frozen) parameters, or vice-versa,
\newline
(ii) in dynamics the $\{S\}$ are the `fast' variables and the 
$\{J\}$ are `slow' variables, or vice-versa, where `fast' and 
`slow' refer to the characteristic microscopic time-scales, 
\newline
(iii) in both cases, the $X$ are intensive control parameters, 
influencing the system deterministically, quenched-randomly  or 
stochastically, 
\newline and  
\newline
(iv) we shall be particularly interested 
in typical behaviour in situations in which any quenched 
disorder is drawn 
independently from identical intensive distributions, enabling 
(at least in principle) useful thermodynamic-limit measures 
of the macroscopic behaviour.

The interest arises when the effects of different interactions 
are `frustrated', in competition with one another. 
In such cases with detailed balance, at low enough noise 
the macrostate structure/space 
is typically fractured (or clustered), in a manner 
often envisaged in terms 
of a `rugged landscape' paradigm in which the 
dynamics is imagined as motion in a very high dimensional 
landscape of exponentially many hills and valleys, often 
hierarchically structured, with concomitant confinements, 
slow dynamics and history dependence. In dynamical systems 
without detailed balance, strictly there is no such simple 
Lyapunov `landscape' 
but the `motion' is analogously complexly hindered, 
with many effective macroscopic time-scales.  

First studied (in physics) in the context of magnetic 
alloys, such systems are now recognised in many different contexts; 
in inanimate physical systems, computer science, and information 
science; in animate biology, economics and social science. 
In these different systems `controllers' of the `control functions'
vary; including the laws of physics, devisors of computer 
algorithms, human behaviour, governmentally-devised laws etc.

\section{The Sherrington-Kirkpatrick model}

A simply-formulated but richly-behaved canonical model is 
that of Sherrington and Kirkpatrick (SK) \cite{SK}, originally 
introduced as a potentially soluble model corresponding to 
a novel mean-field theory 
introduced by Edwards and Anderson 
(EA) \cite{EA} to capture the essential physics of some unusual
magnetic alloys, known as spin glasses \cite{Mydosh, Sherrington07}.

The SK model is 
characterized by a Hamiltonian
\begin{equation}
H=-\sum_{(ij)}J_{ij}\sigma_{i}\sigma_{j}; \sigma =\pm 1
\label{eq:SK}
\end{equation}
where the $i,j$ label spins ${\sigma }$, taken for simplicity 
as Ising, and the interactions $\{J_{ij}\}$ 
are chosen randomly and independently from a distribution $P_{exch}(J_{ij})$. 
Dynamically the system can be considered to follow any standard 
single-spin-flip 
dynamics corresponding to a temperature $T$. Were normal equilibration 
to occur it would be characterized by Boltzmann-Gibbs statistics, $p({\sigma}) 
\sim \exp (-H\{\sigma\}/T)$. 

However, if the distribution $P_{exch}(J)$ has sufficient variance compared 
with its mean and the temperature is sufficiently low,  normal 
equilibration does not occur and complex macro-behaviour results beneath a 
transition temperature.
The interesting regime, known as the `spin glass phase', occurs at intensive 
$T$ if the variance of 
$P_{exch}(J)$ scales with $N$ as $J^{2}/N$, the mean as $J_{0}/N$. 
As Parisi showed, in a series of papers 
({\it e.g.} \cite{Parisi80,Parisi83,MPSTV}) 
which involved amazing insight and highly original 
conceptualization 
and methodology, this glassy state is 
characterized by a hierarchy of `metastable' macrostates, 
differences between restricted and Gibbsian thermodynamic 
averages, as well as non-self-averaging; see 
also {\it e.g.} \cite{MPV}.

These features can be 
characterized by the macrostate overlap distribution functions 
$P(q)=\sum_{S,S'}\delta (q-{\mid{N^{-1}\sum_{i}\langle \sigma_{i} 
{\rangle}_{S}\langle \sigma_{i} {\rangle}_{S'}}\mid)}$ 
where $\langle O {\rangle}_{S}$ denotes a thermodynamic average 
of $O$ over the macrostate $S$. For a conventional system, 
with a single macrostate,
$P(q)$ 
has a single delta function, 
while for a system with entropically extensively many macrostates 
$P(q)$ has more structure.
When the state structure is continuously hierarchical, as it is for SK, 
there is a continuum of weight in the disorder-averaged overlap 
distribution function 
${\bar{P}}(q) =\int {DJ}[ \prod_{(ij)} P_{exch}({J_{ij}})] P_{\{J_{ij}\}}(q)$. 
Non-self-averaging arises 
in different
$P_{\{J_{ij}\}}(q)$ for different realizations of quenched disorder even when 
that disorder 
is chosen {\it i.i.d.} from an intensive $P_{exch}(J)$. 
Ultrametricity \cite{MPSTV} is a feature of the hierarchical order.

Later studies \cite{CuKu}
have further exposed the existence of 
slow dynamics and aging and a remarkable non-trivial quantitative 
relationship to the 
thermodynamics \cite{Young, Parisi2004}. One feature of this is a modification of 
the normal fluctuation-dissipation relationship to $-dR/dC=\beta X(C)$ 
where $R$ is a response function and $C$ a related correlation 
function, parenthetically connected through the measurement 
time $t$ in the limits of large initial 
waiting time/field application time $t_w$ and $t$ itself, 
and $X(C)$ is related to the average 
overlap distribution $\bar P(q)$ by 
$X(C)=\int_{0}^{C} \bar P(q)dq$.
In the normal fluctuation-dissipation theorem $X(C)$
is replaced by unity.  

The original exposure of the subtleties of the SK model
utilised unusual and non-rigorous mathematics and ans\"atze, 
together with unconventional physical conceptualization, going far
beyond the conventional realms of rigorous mathematical physics and
probability theory. The predictions have however long been shown to be
in accord with computer simulations of the model 
({\it e.g.} \cite{Young83})
and consequently were believed by physicists. Their 
rigorous demonstration, though,  
has been a non-trivial challenge which has required deep 
analysis and led to new rigorous mathematical methodologies 
in recent years \cite{Guerra, 
Talagrand98, Talagrand_book, Bovier}, finally completely 
vindicating Parisi's 
theory \cite{Talagrand06}. Thus 
there is no 
doubt that there is deep complexity in the SK model.

On the other hand, while it is generally believed that real
spin glass transitions do occur also in the short-ranged
EA model for dimensions three or more and also in the experimental 
magnetic alloys that first stimulated 
its study, it remains controversial as 
to whether or to what extent all the subtle 
predictions of the SK model apply to these systems 
(see {\it e.g.} \cite{Young}\cite{CK08}). 

Hence it becomes 
appropriate to ask whether range-free frustrated and disordered systems
have a wider relevance beyond as over-idealized models of real magnetic 
alloys and as 
challenges for mathematicians. Possibly remarkably, it turns out
that the answer is a resounding ``Yes''; they turn out to be rather 
ubiquitous in many areas of science.

\section{Beyond magnetic alloys}

Range-free frustrated and disordered many-body problems occur in many 
scenarios outside of physics, for example in many of the hard 
optimization problems studied by computer scientists, and in situations 
in which correlation between individuals occurs through the transfer of 
information available to all, irrespective of physical separation, as 
epitomised by modern interaction through the internet and 
telephones, or commonly available through the world-wide web, newspapers, radio and 
television. These systems 
are usually different in detail from the SK model but share some of the 
same conceptual and technical challenges, as well as providing further 
challenges of their own.

\subsection{The SK model as an optimization exercise}

A simple illustration of the the possibilities of 
extension comes from viewing the SK model 
as an optimization problem which is describable in everyday terms as 
``The Dean's Problem" \cite{Clay}. One imagines a 
University Dean who has to place $N$ students in two 
dormitories, but  with the challenge that every pair of students $(i,j)$ 
either 
likes or dislikes one another to an extent $J_{ij}$ \footnote{$J >0$ 
corresponding to `like' and $J<0$  to dislike.}. His problem is to 
choose to which dorm to allocate each student 
so as to ensure the greatest 
satisfaction overall. Labelling the dorm choices by ${\sigma_{i} = \pm 1}$ 
the SK Hamiltonian becomes the `cost function' that the Dean should
minimise, with the $\{\sigma_{i}\}$ in its ground state the optimal 
choice\footnote{If the Dean also has 
to put an equal number of students in each dorm there is an extra constraint 
$\sum {\sigma_{i}}=1$.}.  Allowing the Dean a degree of uncertainty 
in his decision-making provides an analogue of 
`temperature'. 

\subsection{Simulated annealling}

There are many other combinatorial optimization problems that can be viewed as
finding the minimum of a cost function of the form $H(\{J_{ij...k}\},\{S_{i}\},X)$.

In a `simulated annealing' \cite{Kirkpatrick} `spin-flip' computer 
algorithm to minimise the cost function, an annealing `temperature' $T_A$ is 
introduced artificially
via a stochastic probability measure determined by $\exp(-{\delta H}/T_A)$, 
where $\delta H$ 
is the change in the cost function engendered by the flip, so that 
equilibration at $T_A$ would yield the corresponding Boltzmann 
distribution, and $T_A$ 
is gradually reduced to the value of interest (zero for a minimum of $H$, or 
$T$ if that is the stochastic noise of actual uncertainty). Correspondingly,
$T_A$ can be introduced into an effective equilibrium statistical 
mechanics, with Boltzmann weighting $\exp (-H\{S\}/T_{A})$, examined analytically
and the minimum found from 
$H_{min}= \lim_{T_{A}\to 0}\{-T_{A}\ln\sum_{\{S\}}\exp(-H\{S\}/T_{A})\}$

\subsection{$p$-spin spin glass, satisfiability and error-correction}

In the $p$-spin glass model one replaces the binary interaction of SK by 
one involving $p$ spins; 
\begin{equation}
H=-\sum_{(i_{1}i_{2}..i_{p})}J_{i_{1}i_{2}..i_{p}}\sigma_{i_{1}}
\sigma_{i_{2}}...\sigma_{i_{p}},
\label{eq:p-spin}
\end{equation}
with the $J$ again drawn randomly and independently, all from the same
distribution, and then quenched \footnote{In this case variance
scaling as $N^{-(p-1)}$ yields an intensive transition temperaure.}. 
This apparently innocuous extension of the SK model yields new 
behaviour in several ways. Firstly, instead of a continuous 
onset of a hierarchy of different levels of metastability,  
with a growing continuous range of state overlaps, there is 
a discontinuous onset of many orthogonal but otherwise 
equivalent metastates of finite overlap 
order parameter \footnote{These 
differences can be seen in the character of the onset of structure 
in $P(q)$ at $q > 0$
in addition to the main delta-function peak at $q=0$; 
note that this is in contrast to a conventional ferromagnet 
(or antiferromagnet) 
for which the whole delta-function peak, which in the paramagnetic phase
is at $q=0$, would move to a finite value of $q$.
For a continuous transition $P(q)$ develops extra finite weight 
growing continuously from $q=0$, whereas in a discontinuous 
transition $P(q)$ develops by acquiring
weight directly 
at finite $q=q_1 \sim {\cal O} (1)$ 
but with its weight growing continuously. In the case 
of $p=2$ (SK) the (continuous) onset also has 
finite weight for a 
continuous range of $q$ within $0 \leq q \leq q_{1}$. For
SK $q_1$ 
grows continuously with $(T-T_c)$ where $T_c$ is the 
transition temperature. For $p \geq 3$  
the initial onset is discontinuous 
at a single finite $q_1$. }\cite{GM, CS}. Secondly, the dynamical 
transition is no longer at the same temperature as the 
thermodynamic transition but is higher \cite{Crisanti93}.   
Thirdly, there is another lower temperature (continuous) 
thermodynamic transition to a state with
a continuous range of overlap distributions \cite{Gardner}.  

This type of behaviour, of a dynamical transition 
pre-empting a thermodynamic one, both with discontinuous onset 
of overlap, turns out to be rather common in frustrated 
many-body systems\footnote{In general, the symmetry of definiteness of 
SK and EA seems to  be more the exception than the norm.}. 
So too is the lower temperature thermodynamic transition to 
a continuous range of overlap distributions 
\footnote{Exceptions to this lower transition occur for
so-called spherical spins (individual $S_i$ unbounded but 
with $\sum_{i=1,..N} (S_{i})^{2} = N$) or for $p=\infty$.}. 
Consequently the determination of the minimum achievable 
cost function
often has the difficulties associated with the 
full hierarchical character of the SK model at $T=0$.

In fact, many of the problems of interest in computer science are 
effectively range-free but on
random graphs of finite connectivity, in contrast to the 
full connectivity of the SK- and $p$- spin models of eqns.
(\ref{eq:SK}) and (\ref{eq:p-spin}) above. The conceptual 
ideas extend, albeit made more complicated (and currently 
incompletely solved) by the need 
for higher order overlap functions within the full (replica) 
theory of \cite{EA, SK, Parisi80} \footnote{They also 
no longer 
require inverse $N$-scaling to achieve finite 
transition temperatures; the relevant criterion is scaling as 
$z^{-1}$ where $z$ is the graph coordination number.}. 
Dilution, however, has 
also brought to the fore an alternative and highly 
successful computational methodology in the form 
of `survey propagation' \cite{Mez-Zecch}. 

One example of the conceptual transfer of these ideas 
is to random satisfiability problems in computer science, 
both in explaining the existence of satisfiable-unsatisfiable 
(SAT-UNSAT) phase transitions \cite{Kirk-Sel} and in leading to 
the recognition that for random $K(>2)$-SAT, where $K$ refers to the 
length of the individual clauses to be satisfied 
simultaneously, there should be a region of `HARD-SAT' 
separating practically 
satisfiable SAT problems from UNSAT as the constraint density, 
the ratio of the number of constrained clauses to the 
number of variables, is increased \cite{MPZ02}; this ratio can be 
considered as 
playing a role reminiscent of that of the inverse of temperature 
in the $p$-spin model with the transitions analogues of the $p$-spin 
dynamical and thermodynamic transitions \footnote{Random 
$K$-SAT maps to an Ising spin glass model with terms 
of several $p \leq K$.}. In fact, on closer examination, 
random $K$-SAT exhibits an
even richer sequence of phase transions; see {\it e.g.} \cite{K07}. 

It is possible to interpolate between the type of  
behaviour of the $p \ge 3$ 
model and that of the $p=2$ SK model. One way is to add a magnetic field 
$h$ to the $p$-spin glass. This leads to a sequence 
of behaviours as $h$ is increased; for small $h$ it is 
qualitatively as 
described above for the zero-field $p$-spin model, followed at a 
first critical field by the coming together of the 
dynamical and thermodynamic spin glass transitions 
and replacement of the discontinuous onset of non-trivial 
$P(q)$ by a continuous one \cite{Crisanti93}, but still 
with a single delta function onset at non-zero $q$, 
in addition to that at $q=0$, 
and then at a higher critical field by a transition 
to a continuously distributed hierarchy 
of metastable states and overlaps\footnote{ In fact, 
this sequence of events was first recognised in a 
Potts spin glass \cite{Elderfield, GKS}, still 
with two-body interactions but with the (symmetric) Ising 
interaction $\sigma_{i}\sigma_{j}$ replaced by 
a non-symmetric Potts interaction 
$\delta_{s_{i},s_{j}}; s_{i}=1,2,...p$ where $p$ 
is the Potts dimensionality. In this case the 
sequence is from SK at $p=2$ to pure 
$p$-spin-glass-like at $p=4$}. This suggests 
a possible utility in adding an extra 
`effective field' in the computer algorithmic optimization, 
to avoid the dynamical pre-emption of a thermodynamic 
transition.  

\subsection{Interacting agents}
Another interesting class of range-free problems is of systems 
where many `agents', each with 
individual characteristics but with no direct interactions 
between them, behave in a cooperatively complex fashion
by all reacting to common `information'. This common information
acts as an effectuator for correlation between the agents.
Frustration and complexity arise when the 
goals are such that not all can `win'. 

\subsubsection{The Minority Game}
A minimalist model that illustrates this class is the so-called 
`Minority Game'(MG)\cite{Challet_book, Coolen_book}, 
introduced to emulate some features of a stockmarket 
in which players make profits by buying when the price is low
and selling when the price is high. In a simple version of 
this model $N$ agents at each 
time-step $t$ simultaneously make one of two choices, which we shall denote 
${\pm 1}$. Their `objectives' 
are to make the minority choice. They make their choices on the basis
of (i) some `information' $\bf{I(t)}$ commonly available to all, (ii) the 
operation on that information by each agent $i$ of one of a pair of individual 
strategy operators ${\hat{S}}_{i}^{\alpha};\alpha = +,-$, 
with the output
determining the `choice' made, (iii) individual
`point-scores' $p_{i}(t)$ that enable the agents to `decide' which 
of their two strategies
to employ at each step. The strategy pairs are chosen randomly 
and independently at the outset
and thereafter fixed. The information $\bf{I(t)}$ varies at each 
time-step and 
hence so does the outcome of the strategies acting upon it . The space of
the strategies spans the two possible outputs 
equally. In the simplest deterministic version of the game the strategy 
${\hat{S}}_{i}^{\alpha}$
employed by agent $i$ at time $t$
is that labelled by the same sign as $p_{i}(t)$. The points are updated
according to 
\begin{equation}
p_{i}(t+1) = p_{i}(t) - [{\hat{S}}_{i}^{{\rm sign} (p_{i}(t))}({\bf{I(t)}})]A(t)  
\label{eq:point_update}
\end{equation}
where $[{\hat{S}}_{i}^{\alpha}({\bf I})]= \pm 1$ is  the action choice 
of the strategy ${\hat{S}}_{i}^{\alpha}$  acting on the 
information $\bf I$ and and $A(t)$ is the average `choice' over the stategies
actually employed,
\begin{equation}
A(t)= N^{-1}\sum_{j} [{\hat{S}}_{j}^{{\rm sign} (p_{j}(t))}({\bf{I(t)}})];
\label{eq:majority_weight}
\end{equation}
{\it{i.e.}} by increasing the point-score bias for strategies leading to 
minority behaviour.
In the original formulation \cite{Challet-Zhang} the information used 
was the Booolean string indicating the minority choice in the previous 
$m$ time-steps
of play and the $\hat{S}$ were Boolean operators. However, essentially 
similar behaviour is obtained for a system in which 
${\bf I}(t)$ is randomly generated at each time $t$, equally probably 
from the whole 
space of $m$ binaries \cite{Cavagna}. 

The most obviously relevant macroscopic measure in the MG is the volatility, 
the variance of the choices. Computer simulations demonstrated that it has 
scaling behaviour, the volatility per agent versus the the information 
dimension per agent $d=D/N=2^{m}/N$ approaching independence of $N$ 
as the latter is increased, and also has a cusp-like minimum  at a 
critical $d_c$ with behaviour ergodic for
$d>d_c$ but non-ergodic for $d<d_c$ \footnote{In 
the case shown in Fig. 1, of uncorrelated strategies, the 
cusp-like behaviour is most pronounced for a {\it tabula rasa} start. 
For anti-correlated strategies \cite{GS03} there is no cusp for 
{\it tabua rasa} start but the non-ergodicity-onset is clear.}. 
Fig 1 shows this behaviour for  
\begin{figure}
\centerline{\includegraphics[width=15pc]{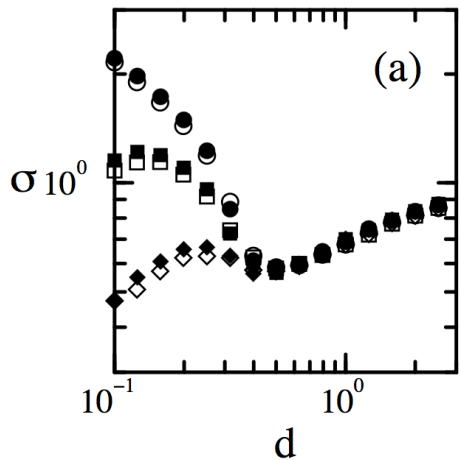}}
\caption{Volatilities in Minority Games with 2 strategies per agent;
Shown are (i) different biases of initial point asymmetries between each 
agent's 2 strategies: $p_i(0)=0.0$ (circles), $0.5$ (squares) and $1.0$ 
(diamonds), 
(ii) a comparison betweeen the results of simulation 
of the deterministic many-agent dynamics (open symbols) and the 
numerical evaluation
of the analytically-derived stochastic single-agent ensemble dynamics. 
From \cite {GS03}.}
\label{figure1}
\end{figure}
a slightly different variant of the model in which the strategies 
are taken as 
$D=dN$-dimensional binary strings ${\bf S}_{i}^{\alpha}=
\{ S_{i}^{{\alpha},1},S_{i}^{{\alpha},2},....S_{i}^{{\alpha},D} \}; i=1,..N, 
\alpha = \pm, $
with each component $S_{i}^{{\alpha},\mu}; \mu =1,..D$ chosen randomly and 
independently at the outset and thereafter fixed (quenched), 
and the stochastic `information' 
consists in randomly choosing $\mu(t)$ at 
each time-step and then using the corresponding strategy elements.
This is reminiscent of the behaviour of 
the susceptibility of the 
SK spin glass, shown in Fig 2,  
\begin{figure}
\centerline{\includegraphics[width=15pc]{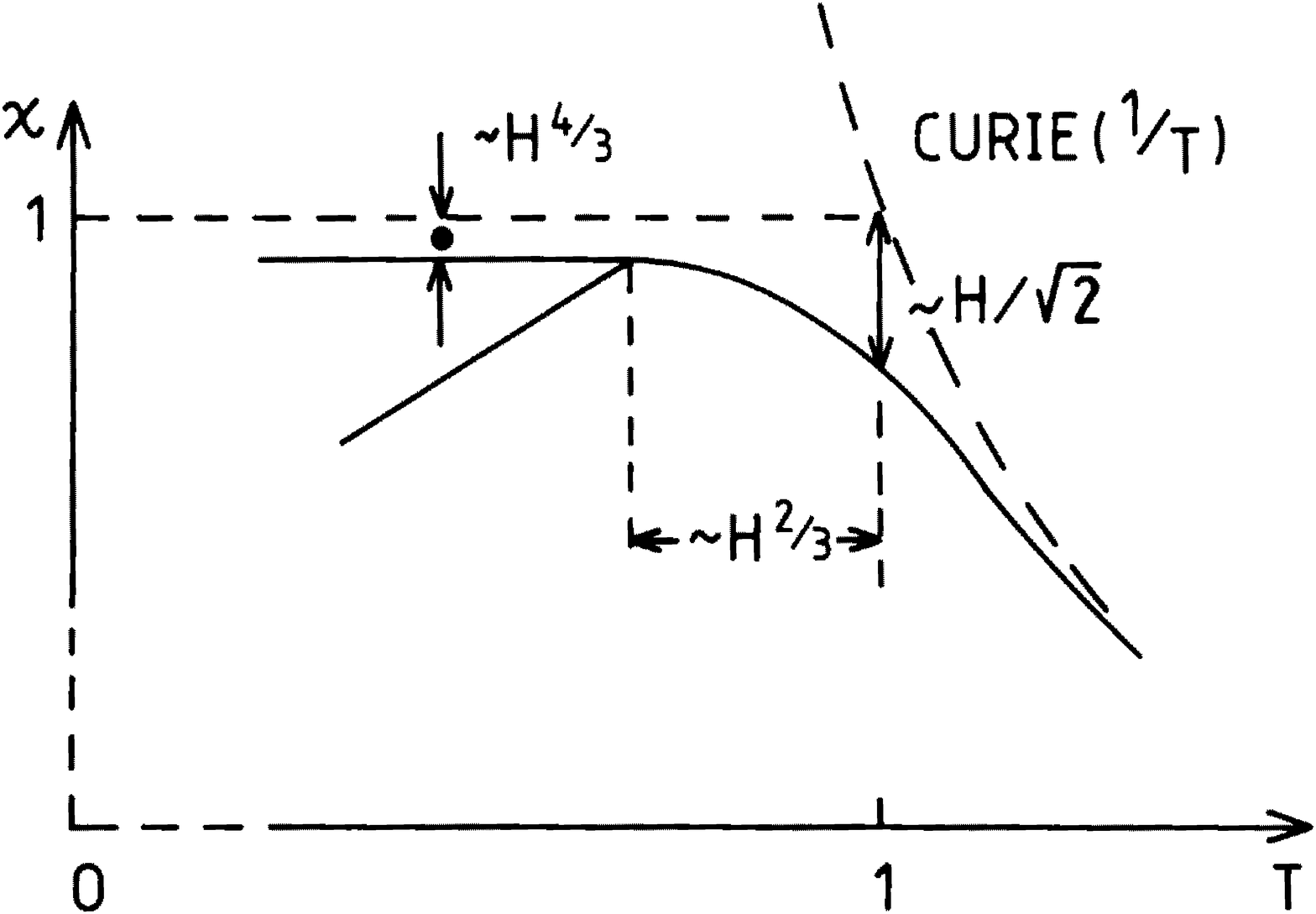}}
\caption{Schematic susceptibility of the SK spin glass in an 
applied field $H$, 
as predicted by Parisi theory.  The upper curve shows the full 
Gibbs average, obtained from the full $q(x)$ and interpreted as 
the field-cooled (FC) susceptibility.  The lower curve shows the 
result of restricting to one thermodynamic state, as obtained from 
$q(1)$ and interpreted as the zero-field-cooled susceptibility.
From  \cite{DS_SG}.}
\label{figure2}
\end{figure}
if one compares the volatility with the 
inverse susceptibility and the information dimension with the 
temperature. Hence one is tempted to analyze the MG using 
methodolgy developed for spin glasses. 

Updating the point-score only after $M$ steps where $M \ge {\cal{O}}(N)$ 
leads to an averaging over the random 
information to produce an effective interaction between the agents and 
yield the so-called `batch' game (with temporally-rescaled update dynamics)
\begin{equation}
%\hspace{-0.5cm}
p_i(t+1)=p_i(t)  -\sum_{j} J_{ij} \mbox{sgn}{(p_j(t))}- 
h_i \equiv p_i(t) - \partial{H}/\partial{s_i}\mid_{s_i = \mbox{sgn}(p_i(t))},
\label{batchupdate}
\end{equation}
where $H$ is an effective `Hamiltonian'
\begin{equation}
H = \sum_{(ij)} J_{ij}s_{i}s_{j} + h_{i} s_{i}
\end{equation}
and $J_{ij}$ and $h_{i}$ are effective `exchange' and 
`field' terms given by
\begin{equation}
J_{ij} = N^{-1}\sum_{\mu=1}^{D}{\xi_{i}^\mu}{\xi_{j}^\mu}, \\\
h_{i} = N^{-1/2}\sum_{\mu=1}^{D}
{\omega_{i}^\mu}{\xi_{i}^\mu},
\end{equation}
where
${\bf \omega}_i=({\bf S}_{i}^{1}+{\bf S}_{i}^{2})/2,\,{
\bf \xi}_i=({\bf S}_{i}^{1}-{\bf S}_{i}^{2})/2$. Since 
the $\bf{S}$ are random so are the exchange and field terms. 
Hence $H$ is a disordered and frustrated control function. The 
expression for the $\{J_{ij}\}$ is very reminiscent of
the Hebbian-inspired synapses of the Hopfield 
neural network model \cite{Hopfield}, where the $\{\xi_{i}^\mu\}$ 
are the stored memories, but crucially with the opposite sign ensuring 
that here the $\{\xi_{i}^\mu \}$ are now repellors rather 
attractors\footnote{Note also that the usual spin glass or
neural network dynamics is different in detail from that of the 
minority game, {\it e.g.} random sequential rather than parallel.}.

\section*{Methodologies}

There are two main methodologies employed to study statics,
the replica procedure and the cavity method (see {\it e.g.} \cite{MPV}).
The most common method for the cooperative dynamics is the 
generating functional 
method \cite{deD, Jan}. 

In the replica method  
one studies the disorder-averaged free energy
\begin{equation} 
\int D\{J\} P_{exch}(\{J\}) (-T\ln {\rm Tr}_{\{\sigma\}} \exp(-H_{\{J\}}
(\{\sigma\})/T)),
\end{equation}
using the identity $\ln Z = {\rm Lim}_{n \to 0}\{ Z^{n} -1\}/n$, 
identifying the power $n$ as describing $n$ replicas, $\alpha = 1,..n$;
with $n$ eventually taken to 0. 
Macroscopic 
order parameters are introduced through multiplication 
by unity of the form
\begin{equation}
1=\int \prod_{(\alpha \beta)} D{q^{\alpha\beta}} \delta(q^{\alpha\beta}-N^{-1}
\sum \langle \sigma_{i}^{\alpha} \sigma_{i}^{\beta} {\rangle}_{H_{eff}}), 
\end{equation}
where the $\alpha,\beta$ label replicas and $H_{eff}$ is the effective 
Hamiltonian after disorder averaging. The 
microscopic variables $\{\sigma_{i}^{\alpha}\}$ are integrated out 
and the
dominant extremum \footnote{The correct extremum is actually the maximum
\cite{MPV, Guerra, Talagrand_book}.} with respect to 
the $q^{\alpha \beta}$ is taken in the limit
$N \to \infty$. In the most natural asantz, replica symmetry among 
$q^{\alpha \beta}; \alpha \neq \beta$ was assumed 
\cite{EA,SK},  but this proved to be too naive. 
The correct solution for the SK model 
requires Parisi's much more subtle ansatz of replica 
symmetry breaking \cite{Parisi80}. This ansatz introduces a hierarchy of
spontaneous replica symmetry breaking (RSB) with a sequence of 
$q_i ,x_i ; i=1,..K$
that in the limit of $K \to \infty$ yields a continuous order function
$q(x): 0 \leq x \leq 1$, later  shown \cite{Parisi83} to be related to the 
average overlap distribution through 
${\bar{P}}(q) = \int{dx} \delta (q - q(x))$.

The dynamical functional method for the SK model is discussed in 
\cite{CuKu,CK08}. Here we describe instead its use for
the Minority Game \cite{Coolen_book}. A generating functional can be defined by
\begin{equation}
Z =\int {\prod_{t}} d{\bf p}(t) W({\bf p}(t+1) \mid {\bf p}(t)) P_0 ({\bf p}(0)), 
\label{generatingfunctional}
\end{equation}
where ${\bf p}(t)=(p_1(t),\dots,p_N(t))$, 
 $W({\bf p}(t+1)\mid {\bf p(t)})$ denotes the transformation 
operatation of eqn. (\ref{batchupdate})
 and  $P_0({\bf p}(0))$ denotes the probability distribution of the 
initial score differences.

Averaging over the specific choices of quenched strategies, introducing 
macroscopic 
two-time correlation order functions via 
\begin{equation}
1=\int \prod_{t,t'} DC(t,t') \delta (C(t,t') -N^{-1}\sum_{i}{\rm sign}p_{i}(t)
{\rm sign}p_{i}(t'))
\end{equation}
\label{unity}
and similar expressions for response functions $G(t,t')$
and derivative-variable correlators $K(t,t')$, and 
integrating out the microscopic variables,
the averaged generating functional may then be transformed exactly into a form 
\begin{equation}
Z=\int D{\bf C}D{{\bf\tilde{C}}} D{\bf G} D{{\bf\tilde{G}}} D{\bf K} 
D{{\bf \tilde{K}}} 
\exp \left(N \Phi({\bf{C}},{\bf{\tilde{C}}}, {\bf{G}}, 
{\bf{\tilde{G}}}, {\bf{K}}, {\bf{\tilde{K}}}\right),
\label{extremalfunctional}
\end{equation}
where $\Phi$ is $N$-independent, the bold-face notation 
denotes matrices in time  and the tilded variables are complementary ones
introduced to exponentiate the delta functions in eqn. (\ref{unity}) 
and its partners. 
Being extremally dominated, in the large-$N$ limit this yields
the effective single agent stochastic dynamics
\begin{equation}
p(t+1) = p(t) -\alpha \sum_{t' \leq t} ({\bf{1}} + {\bf{G}})^{-1}_{tt'} 
\mbox{sgn}p(t') + {\sqrt \alpha} \eta(t),
\label{effectiveagent}
\end{equation}
where $\eta(t)$ is coloured noise determined self-consistently over the 
corresponding ensemble by
\begin{equation}
\langle \eta(t) \eta(t') \rangle = [({\bf{1}} + 
{\bf{G}})^{-1}({\bf{1}}+{\bf{C}})({\bf{1}}+{\bf{G^T}})^{-1}]_{tt'}.
\label{colourednoise}
\end{equation}
Fig. 1 demonstrates the veracity of this result in a comparison of the 
results of computer simulation of the original deterministic many-body 
problem eqn. (11) and the numerical evaluation of the 
self-consistently noisy single-agent ensemble of eqn. (\ref{effectiveagent}).

The analogous equations for the $p$-spin spherical spin glass formed the 
basis for recognition of the dynamical transitions mentioned earlier and 
the existence of aging solutions and modifications to 
conventional fluctuation-dissipation relations.

\section{Critical behaviour and correlation length}

Having commented earlier that standard non-frustrated non-disordered 
infinite-ranged systems do not have interesting critical behaviour, 
it is relevant to note that again frustrated disordered systems are 
different \cite{Opper, OSS07, OS08_2, Pankov}, having interesting 
critical behaviour at low temperature and applied magnetic field, 
even though mean-field. 

Parisi replica symmetry breaking involves an infinite sequence of 
hierarchies. $K$-RSB has $K$ step-breaks in the order function 
$q(x): {0 \leq x \leq 1}$ \footnote{$q(x)$ is  related to the 
averaged overlap distribution $\bar{P}(q)$
by ${\bar{P}}(q) = \int{dx} \delta (q - q(x))$.}. The exact 
free energy is formally obtained by finding the supremum with 
respect to the break and plateau values $q_{i},x_{i}$ and 
taking $K \to \infty$ \cite{Parisi80,Talagrand06}. The continuum 
limit was given as a set of implicit equations already in 
Parisi's early work \footnote{See also \cite{SD}.}. Most 
(but not all) of the subsequent 
analysis has 
been perturbative near to the transition temperature for spin 
glass onset\footnote{For the most complete perturbative
study the reader is referred to \cite{CR}.}. Numerical 
evaluations have until a couple of years ago
been restricted to just the first few steps of RSB, but 
very recently very high accuracy numerical 
extremizations for high orders of RSB have been performed at 
zero and low temperatures and have shown interesting features 
\cite{Opper, OSS07,OS08_1,OS08_2}.

At low temperatures the steps $x_{i}$ scale as $x_{i} \sim a_{i}T$ 
with the $a_{i}$ having non-zero limits as $T \to 0$ and 
exposing critical points at both $a =0$ and $a = \infty$. 
As $T \to \infty$ the $K$-step approximation of ${q_{i}}$ 
against $a_{i}$ approaches a fixed-point function $q{^{*}}(a)$ 
of form close to $q^{*}(a) = ({\sqrt \pi} / {2a}){\rm {erf}} (\xi /a)$ 
with $\xi$ a `correlation function' in $a$-space given by 
$\xi \approx 2/\sqrt \pi$ 
\footnote{Strictly the behaviour is found to deviate 
slightly but subtly - see Refs. \cite{OSS07, OS08_2}.}. 
The degree of RSB can be viewed as an effective 
one-dimensional lattice of size $K$, with $K \to 
\infty$ the analogue of the infinite-length lattice 
(or thermodynamic) limit. Similarly, 
finite-$K$ approximation yields an analogue of finite-size 
effects, including finite-size scaling. Note 
however that this new type of finite-size scaling is 
for a mean-field 
problem in the thermodynamic limit and is in a space of 
degree of approximation. There are also finite $K$-size scalings 
when the system is perturbed away from the $T=0$ critical
point (at $a=0$) and for finite applied field $h$ near 
$a=\infty$ \cite{OS08_2}. Correspondingly there are 
further `correlation lengths' in temperature-deviation 
and in field-deviation, which of course also determine the 
extent of RSB needed to get a good approximation as 
temperature or field become non-zero.

\section*{Conclusions}

In this short paper it has only been possible to present a 
brief and non-detailed vignette of the complexity that can and 
does exist in disordered and frustrated many-body systems, even 
within a dimension-free mean-field situation. The puzzles, 
intrigues and challenges have developed and been a source
of intense study for over 30 years. Finite-range systems have also been a 
great source of interest, again with significant progress but 
still subject to some controversy \cite{Young08,Parisi08}.

The case of systems with variables having different fundamental
timescales, such as fast neurons and slow synapses or evolutionary 
models with different timescales for phenotypes and genotypes, have 
not been discussed. Nor has the problem of dynamical sticking in 
effectively self-determined disordered states of some systems 
without quenched disorder in their control functions but
started far from equilibrium. 

Also, in this brief review, only some of the simplest models 
have been described. It is however clear that many extensions 
and more realistic/complete scenarios  
exist that are still effectively range-free, yet complex, 
interesting
and challenging.

\section*{Acknowledgements}
The author would like to thank his numerous collaborators, 
students, colleagues and friends, too many to name all individually, 
for their parts in helping his understanding and appreciation of 
the subject of this paper. He also acknowledges, with gratitude, 
the financial support of the EPSRC (and its predecessors), the 
EC and the ESF.

%\section*{References}
%\vspace{-2.5cm}

  %

\begin{thebibliography}{99}
\bibitem{Cardy} J Cardy, {\it Scaling and Renormalization 
in Statistical Physics}
(Cambridge University Press, Cambridge, 1996)
\bibitem{SK} D. Sherrington and S. Kirkpatrick, 
%{\it Solvable Model of a Spin-Glass},
  {\em Phys. Rev. Lett.}{\bf 35} 1972 (1976)
\bibitem{EA} S.F. Edwards  and P. W. Anderson, 
%{\it Theory of spin glasses}, 
{\em J. Phys. F} {\bf{5}}, 965 (1975)
\bibitem{Mydosh} J. A. Mydosh, {\it Spin glasses; an 
experimental introduction}, Taylor and Francis, London (1993)
\bibitem{Sherrington07} D. Sherrington, in {\it Spin Glasses}, 
eds. E. Bolthausen and A. Bovier, (Springer, Berlin, 2007) 
\bibitem{Parisi80} G. Parisi, 
%{\it The order parameter for spin glasses: A function on the interval 0-1}, 
{\em J. Phys. A} {\bf 13}, 1101 (1980) 
\bibitem{Parisi83} G. Parisi, 
%{\it Order Parameter for Spin-Glasses}, 
{\em Phys. Rev. Lett.} {\bf 50}, 1946 (1983)
\bibitem{MPSTV}
M. M\'ezard, G. Parisi, N. Sourlas, G. Toulouse  and M.A. Virasoro, 
%{\it Replica symmetry breaking and the nature of the spin glass phase}
{\em J.Physique} {\bf{45}}, 843 (1984) 
\bibitem{MPV} M. M\'ezard, G. Parisi and M.A. Virasoro,  
{\it Spin GlassTheory and Beyond} (World-Scientific, Singapore, 1987)
\bibitem{CuKu}
L. F. Cugliandolo and J. Kurchan, 
%{\it On the out-of-equilibrium relaxationof the Sherrington-Kirkpatrick model}, 
{\em J.Phys.A} {\bf {27}}, 5749 (1993)
\bibitem{Young} A.P. Young A P (ed.) {\it Spin Glasses and Random Fields} 
(World Scientific, Singapore, 1997)  
\bibitem{Parisi2004} G. Parisi, 
%{\it The overlap in glassy systems}, 
in 
{\it Stealing the Gold: a Celebration of 
the Pioneering Physics of Sam
Edwards},  eds. P. M. Goldbart, N. Goldenfeld and 
D. Sherrington (Oxford University Press, Oxford, 2004)
\bibitem{Young83} A. P. Young, 
%{\it Direct Determination of the Probability Distribution for 
% the Spin-glass Order Parameter}, 
{\em Phys. Rev. Lett.} {\bf 51}, 1206 (1983)
\bibitem{Talagrand98} M.Talagrand, 
{\it The Sherrington-Kirkpatrick model: a challenge for mathematicians},
{\em Probab. Theor. Rel.} {\bf 110}, 109 (1998)
\bibitem{Talagrand_book} M. Talagrand {\it Spin Glasses: 
a Challenge for Mathematicians} 
(Springer, Berlin, 2003)
\bibitem{Guerra} F. Guerra, 
% {\it Spin Glasses}, 
{\em cond-mat/}057581 (2005)
\bibitem{Bovier} E. Bolthausen and A. Bovier eds., 
{\it Spin Glasses} (Springer, Berlin, 2007)
\bibitem{Talagrand06} M. Talagrand, 
% {\it The Parisi formula}, 
{\em Ann. Math.}{\bf 163}, 221 (2006)
\bibitem{CK08} L. F. Cugliandolo and J. Kurchan, 
%{\it The out of equilibrium dynamics of the Sherrington-Kirkpatrick model}, 
{\em J. Phys. A}{\bf 41}, 324018 (2008)
\bibitem{Clay} A variant was posed to describe one of the 
Clay Millenium Prize Problems; see http://www.claymath.org/millennium/P\_vs\_NP. 
That the problem of finding the ground state of a spin glass in three 
and more
dimensions is NP-complete has 
been known since at least 
the early 1980s. 
\bibitem{Kirkpatrick} S. Kirkpatrick, C. D. Gelatt and M. P. Vecchi, 
%{\it Optimization by Simulated Annealing}, 
{\em Science} {\bf 220}, 672 (1983)
\bibitem{Gross} D. J. Gross, I. Kanter and H. Sompolinsky, 
%{\it Mean field theory of the Potts glass}
{\em Phys.Rev.Lett.}{\bf 55}, 304 (185)
\bibitem{Kirkpatrick_W} T. Kirkpatrick and P. G. Wolynes, 
%{\Stable and metastable states in mean-field Potts and structural glasses}
{\em Phys.Rev.}{\bf B} 36, 8552 (1987)
\bibitem{Mez-Zecch} M. M\'ezard and R. Zecchina, 
{\em Phys.Rev. E}{\bf 66}, 056126 (2002)
\bibitem{GM} D. J. Gross and M. M\'ezard, {\em Nuc. Phys. B}{\bf 240}, 431 (1984)
\bibitem{CS} A. Crisanti and H-J Sommers, {\em Z. Phys. B}{\bf 87}, 341 (1992) 
\bibitem{GNS} P. Gillin, H. Nishimori and D. Sherrington, {\em J. Phys. A}{\bf 34}, 2949 (2001)
\bibitem{MPZ02} M. M\'ezard, G. Parisi and R. Zecchina, {\em Science} {\bf 297}, 
812 (2002) 
\bibitem{K07} F. Kzakala, A. Montanari. F. Ricci-Tersenghi, G. Semerjian and L. Zdeborova,
{\em Proc. Nat. Acad. Sci.} {\bf 104}, 10318 (2007)
\bibitem{Crisanti93} A. Crisanti, H. Horner and H-J. Sommers, 
{\em Z. Phys. B},{\bf 92}, 257 (1993) 
\bibitem{Gardner} E. Gardner, {\em Nuc. Phys. B} {\bf 257}, 747 (1985)
\bibitem{Kirk-Sel} S. Kirkpatrick amd B. Selman, {\em Science} {\bf 264}, 
1297 (1994)
\bibitem{Elderfield} D. Elderfield and D. Sherrington, {\em J. Phys. C}{\bf 16}, 
L497 (1983)
\bibitem{GKS} D. J. Gross, I Kanter and H. Sompolinsky, {\em Phys. Rev. Lett.}
{\bf 55}, 304 (1985)
\bibitem{Challet_book} D. Challet, M. Marsili and Y-C Zhang, {\it Minority Games}
(Oxford University Press, Oxford 2005) 
\bibitem{Coolen_book} A.C.C. Coolen, {\it The Mathematical Theory of Minority Games} (Oxford University Press, Oxford 2005)
\bibitem{Challet-Zhang} D. Challet and Y-C Zhang, {\em Physica A}{\bf 246}, 407
(1997)
\bibitem{Cavagna} A. Cavagna, {\em Phys. Rev E}{\bf 59}, R3783 (1998)
\bibitem{GS03} T.Galla and D. Sherrington {\em Physica A} 324, 25 (2003) 
\bibitem{DS_SG} D. Sherrington, in {\it Heidelberg Symposium on Glassy Dynamics}, 2, (Springer-Verlag, Berlin 1987)
\bibitem{Hopfield} J. J. Hopfield, {\em Proc.Nat.Acad.USA}{\bf 79}, 2554 (1982)
\bibitem{deD} C. de Dominicis, {\em. J. Physique C}{\bf 1}, 247 (1976)
\bibitem{Jan} H.Janssen,{\em Z. Phys. B}{\bf 23}, 377 (1976)
\bibitem{Young08} A. P. Young, {\em J.Phys.A}{41},324016 (2008)
\bibitem{Parisi08} G. Parisi, {\em J. Phys. A.}{\bf 41}, 324002 (2008)
\bibitem{Opper} R. Oppermann and D. Sherrington, {\em Phys. Rev. Lett.} 
{\bf 95}, 197203 (2005)
\bibitem{OSS07} R. Oppermann, M. J. Schmidt and D. Sherrington, 
{\em Phys. Rev. Lett.}{\bf 98},127201 (2007)
\bibitem{OS08_1} R.Oppermann and M. J. Schmidt, {\em arXiv:} o801.1756 (2008)
\bibitem{OS08_2} R. Oppermann and M. J. Schmidt, {\em arXiv:}0803.3918 (2008)
\bibitem{Pankov} S. Pankov, {\em Phys. Rev. Lett.}{\bf 96}, 197204 (2006)
\bibitem{SD} H-J. Sommers and W. Dupont, {\em J. Phys. F}{\bf 17}, 5785 (1984)
\bibitem{CR} A. Crisanti and T. Rizzo, {\em Phys. Rev. E}{\bf 65}, 
046137 (2002)
\end{thebibliography}
\end{document}